\documentstyle[12pt]{article}

\let\na=\nabla

\let\a=\alpha
\let\b=\beta
\let\g=\gamma
\let\d=\delta

\let\h=\eta

\let\l=\lambda
\let\m=\mu
\let\n=\nu

\let\r=\rho
\let\s=\sigma
\let\t=\tau

\def\2{{1\over2}} \def\4{{1\over4}} \def\52{{5\over2}}
\def\6{\partial }

\def\({\left(} \def\){\right)} \def\<{\langle } \def\>{\rangle }

\def\beg{\begin{equation}}
\def\begar{\begin{eqnarray}}
\def\ee{\end{equation}}
\def\ea{\end{eqnarray}}

\newcommand{\pref}[1]{(\ref{#1})}                % Rererenz auf Label
\newcommand{\plabel}[1]{\label{#1}}              % Label errichten
\newcommand{\pcite}[1]{\cite{#1}}                % Zitat aus Bibliographie
\newcommand{\pbib}[1]{\bibitem{#1}}              % Biblio-Eintragung

\def\nh{\hat{\nabla}}
\def\gh{\hat{g}}
\begin{document}

\begin{titlepage}
\hfill TUW-93-19
 \begin{center}
\Large
{\bf On the definiteness of the conformal anomaly in
   nonconformal gauges}\\[1cm]
\normalsize
 W. M\"odritsch\\[1.2cm]
 Institut f\"ur Theoretische Physik, TU-Wien\\
 1040 Wien, Wiedner Hauptstra\ss e 8-10/136, Austria\\
 FAX: (+43-1) 567760, e-mail: wmoedrit@ecxph.tuwien.ac.at\\
\end{center}
\vspace{1.2cm}

\centerline{{\bf Abstract}}
The critical dimension of the bosonic string in the harmonic and the
deDonder gauge may be calculated from the time ordered product
of two energy momentum tensors. We show that recently found ambiguities
within that method in nonconformal gauges can be resolved by a
treatment respecting background covariance.

\vspace{2.5cm}
\end{titlepage}

\newpage

The well-known central feature of the bosonic string is
that it can only consistenly be formulated in 26 dimensions.
While this result is well established in the usual conformal gauge, there
are only a few calculations verifying this result in nonconformal gauges
\pcite{Duese,RK,KR,Baul}.
However, it was claimed recently \pcite{Abe} that due to ambiguities
in the energy momentum tensor (e-m tensor) in those gauges, it may
happen that the coefficient of the conformal anomaly cannot
be computed in a unique way. This would be a serious contradiction
to existing theorems which prove the gauge-independence of anomalies,
even including anomalies of global symmetries in general gauge theories
\pcite{Pigu}, or to a theorem which states that the anomalous
vertex is the time ordered product of two e-m tensors \pcite{Baul}.
According to those theorems, anomalies
in different gauges can only differ by terms which may be absorbed as
counter terms to the action.

It is the purpose of this note to show that by using a uniquely defined
background covariant e-m tensor the usual result is obtained, also with the
method of \pcite{Abe}. Two different (nonconformal) gauges are studied
for this purpose.

To the free bosonic string action $(g=-det g_{\a \b})$
\beg
 S_x = \2 \int d^2x \sqrt{g} g^{\a \b} \6_{\a} X^M \6_{\b} X_M
\ee
we add a Faddeev-Popov and gauge fixing term in order that diffeomorphism
and also Weyl rescalings are fixed completely:
\beg  \plabel{fpdiff}
  S_{\phi \pi}+S_{gf} = i \int d^2x \,\, s[\bar{c}_{\a} F^{\a}(g,\hat{g}) +
       \bar{\t} F(g,\hat{g})]
\ee
\newpage
We have introduced the nilpotent BRST operator $s$:
\begar
&&s g^{\a \b} = -\6_{\l} c^{\a} g^{\l \b} -\6_{\l} c^{\b} g^{\l \a}+
                 c^{\l} \6_{\l} g^{\a \b} -\t  g^{\a \b} \nonumber \\
&&s c^{\a} = c^{\l} \6_{\l} c^{\a} , \qquad \qquad s\t = c^{\l} \6_{\l} \t
             \nonumber \\
&&s X^M = c^{\l} \6_{\l} X^{M}       \plabel{BRST}\\
&&s \bar{c}_{\a} = i B_{\a}, \qquad \qquad s \bar{\t} =  i B   \nonumber
\ea

The harmonic gauge is defined by the gauge fixing conditions
(cf. \pcite{Duese,RK,KR})
\begar
         F^{\a}(g,\hat{g}) &=&  \hat{\na}_{\b} \sqrt{g} g^{\a \b}\\
         F(g,\hat{g}) &=& \sqrt{\hat{g}} \hat{g}^{\a \b} h_{\a \b},
         \plabel{trace}
\ea
where we have choosen the gauge fixing metric to be equal to the classical
part of $g_{\a \b}$, which need not be the case in general. This means
that
\beg
      h_{\a \b} = g_{\a \b}-\hat{g}_{\a \b}
\ee
is our quantum field.
As shown in \pcite{KR} it is possible to linearize the action
with respect to $h$ without changing the physics because only one-loop
graphs contribute since $h$ does not propagate - except by changing into
$B$. Furthermore eq. \pref{trace}
leads to an algebraic equation for $h$ and can integrated out immediately.
This means that $h$ has to be considered as traceless with respect to
$\hat{g}$ in the following. For the calculation of the time ordered
product of two e-m tensors to one loop order we only need the part from
the action which is bilinear in the fields. Then the diffeomorphism ghost
and gauge fixing part \pref{fpdiff} becomes
\beg
  S_{\phi \pi}+S_{gf} = \int d^2x \sqrt{\hat{g}}\, [ i \bar{c}_{\a}
   (\hat{\na}_{\g} \hat{\na}^{\g} -\2 \hat{R}) c^{\a}  + B_{\a}
   \hat{\na}_{\b} h^{\a \b}] \plabel{Sgf}
\ee
The e-m tensor
\beg
 T_{\a \b} = \frac{2}{\sqrt{\hat{g}}} \frac{\d S}{\d \hat{g}^{\a \b}}
 \plabel{Tdef}
\ee
is by definition conserved with respect to the
background covariant derivative. \pref{Tdef} will be discussed further
below.
With eqs. \pref{Sgf} and  \pref{Tdef} the
ghost and $B-h$ parts of the e-m tensor are obtained in an unambiguous
way.
\begar
 T_{\a \b}^{\bar{c}c} &=& 2i [ \hat{\na}_{( \a} \bar{c}^{\g}
 \hat{\na}_{\b )} c_{\g} -
        \hat{\na}^{\g} \bar{c}_{( \a} \hat{\na}_{\g} c_{\b )} +
        \hat{\na}_{\g} (\bar{c}^{\g} \hat{\na}_{( \a} c_{\b )} +
                        \hat{\na}_{( \a} \bar{c}_{\b )} c^{\g} )-
                        \nonumber \\
      & & \qquad - \frac{g_{\a \b}}{2} ( \hat{\na}^{\g} \bar{c}^{\s}
      \hat{\na}_{\g} c_{\s}+
             \hat{\na}_{\g} \hat{\na}_{\s} ( \bar{c}^{\s} c^{\g})+
             \bar{c}^{\s} R_{\s \g} c^{\g}) -  \plabel{Tcc} \\
      & & \qquad - \bar{c}_{( \a} (\hat{\na}_{\g} \hat{\na}^{\g} -\2
      \hat{R}) c_{\b )} ] \nonumber\\
 T_{\a \b}^{B h} &=& -2 (\hat{\na}_{( \a} B^{\g} ) h_{\b ) \g} - B^{\g}
 \hat{\na}_{ \g} h_{\a \b} +
                     \hat{g}_{\a \b} \hat{g}^{\m \n} ( \hat{\na}_{\m}
                     B^{\g} h_{\n \g} )
                     +2 B_{(\a} \hat{\na}^{\s} h_{\s \b )}  \plabel{TBh}
\ea
The last terms of these e-m tensors are essentially equations of motion
and could have been avoided, if we had decided to take $\bar{c}^{\a}$
and $B^{\a}$ as $\hat{g}$ independent fields, instead of $\bar{c}_{\a}$
and $B_{\a}$.
They do not contribute to the anomaly.

The computation of the conformal anomaly starts from the flat limit
of $\gh$ in \pref{Tcc} and \pref{TBh}. Before stating the result of this
calculation, we turn to the method used in \pcite{Abe}. There $\gh$ was
choosen to be flat already at the level of the action.  Ambiguities
may arise in this case because by covariantizing ordinary derivatives,
ordering problems appear. Consider the term with $\hat{R}$ in \pref{Sgf}.
It obviously vanishes in the flat limit, but its contribution to the
e-m tensor is readily calculated as

%\newpage
\begar
   \lim_{\gh \to \h}  T_{\a \b}^R &:=& -\lim_{\gh \to \h} i \frac{2}{
   \sqrt{\hat{g}}} \frac{\d}{\d \gh^{\a \b}}
          \int d^2x \sqrt{\gh} \bar{c}^{\m} \hat{R}_{\m \n} c^{\n} =
          \plabel{TR} \\
             &=& -\lim_{\gh \to \h} i [ \nh_{\r} \nh_{(\a} ( \bar{c}_{\b )}
             c^{\r}) +
                     \nh_{\r} \nh_{(\a} ( \bar{c}^{\r} c_{\b )})-
                     \nh_{\r} \nh^{\r} ( \bar{c}_{(\a} c^{\b)})-
                     \nonumber \\
             & &  \qquad - \gh_{\a \b} (\nh_{\m} \nh_{\n} ( \bar{c}^{\m}
             c^{\n})+
                 \bar{c}^{\m} \hat{R}_{\m \n} c^{\n}) + \hat{R}
                 \bar{c}_{(\a} c_{\b)}] = \nonumber \\
            &=& -i \{ \6_{\s} [ \6_{(\a}( \bar{c}^{\s} c_{\b )} +
            \bar{c}_{\b )} c^{\s})-
                  \6^{\s} (\bar{c}_{(\a} c_{\b )})] - \h_{\a \b} \6_{\s}
                  \6_{\r} (\bar{c}^{\s} c^{\r}) \}
                   \nonumber
\ea
This turns out to exactly coincide with the ambiguous term of ref
\pcite{Abe}.
Thus, in a covariant approach, this term must not be traced back to a
total derivative in the action, but to the curvature term above.
It should be stressed that the factor in front of this term is completely
determined by the gauge-choice which led to \pref{Sgf} so that no
ambiguity remains if the limit $\gh \to \eta$ is taken \underline{after}
computing \pref{Tcc} and \pref{TBh} and not before.

We also note that \pref{TR} is symmetric and conserved without the
equations of motion in the flat limit
\beg
 \6^{\a} \lim_{\gh \to \h}  T_{\a \b}^R = 0.
\ee
This situation appears to be the analogue of the well known fact that
in the conformal gauge one can add to the flat e-m tensor the quantity
\beg \plabel{tpp}
  t_{++} = \6_{+}(c^{+} b_{++}),
\ee
which is conserved by virtue of the equations of motion.
The coefficient of this quantity is also only fixed by performing
the calculation on the curved world sheet. Furthermore, Eq. \pref{tpp} can
be derived also from a curvature term if the ghost action is formulated
in the bosonized language \pcite{GSW}.

In our fully background covariant approach we now calculate the anomaly
from the flat limit of the e-m tensor \pref{Tcc}+\pref{TBh}, including the
usual part from the $X$ fields  by considering
the nonlocal pieces of the time ordered product of two e-m tensors.
The result is
\beg
 \< T T_{\a \b}(x) T_{\g \d}(y) \>  = \frac{i c}{12 \pi} \int d^2k e^{i k
    (x-y)}
         \frac{k_{\a} k_{\b} k_{\g} k_{\d}}{k^2} + \mbox{local terms}
         \plabel{TTT}
\ee
with the constant $c$ given in tab. 1 where $X, c\bar{c}$, and $Bh$ denote
the respective contribution of the loops with boson, F.P. ghost and
with $B-h$ propagators.
The fact that the time ordered product of two e-m tensors as defined by
eq. \pref{Tdef} is related to
the anomalous vertex can also be understood from another point of view. If
one recognizes
the effective action as the underlying theoretical concept, one
can show \pcite{Baul} that the second derivative of the effective
action with respect to the metric $\gh$ (the classical part of $g$)
is the anomalous vertex. Since the nonlocal part of that
quantity is equal to the nonlocal part of eq. \pref{TTT}, if
the e-m tensor is defined as in eq. \pref{Tdef}, this argument
shows that we have calculated the usual conformal anomaly.

As a further illustration of the anomaly calculation in non-conformal
gauges in the operator formalism we choose the deDonder gauge:
\begar
         F^{\a}(g,\hat{g}) &=&  \6_{\b}(\gh^{\a \b}-g^{\a \b}) = \6_{\b}
         h^{\a \b} \\
         F(g,\hat{g}) &=& \hat{g}^{\a \b} h_{\a \b} \plabel{trace2}
\ea
The linearized action of the ghost and auxiliary fields (without
interaction terms and with the Weyl ghosts integrated out) reads:
\beg
  S_{\phi \pi}+S_{gf} = \int d^2x  [ i \bar{c}_{\b} \6_{\a}
   (2 \hat{\na}^{( \a} c^{\b )} - \gh^{\a \b} \nh_{\r} c^{\r}) -
       B_{\b} \6_{\a} h^{\a \b} ] \plabel{Sgf2}
\ee
It should be mentioned that the flat limit of this part of the
action is the same as that of eq. \pref{Sgf}. However, the
two e-m tensors differ. In the approach of refs. \pcite{RK,KR} and
\pcite{Baul} this is reflected by the fact that the vertices
of the background field depend on the gauge.

The e-m tensor derived with \pref{Tdef} from the
action \pref{Sgf2} reads
\begar
 T^{\bar{c}c}_{\a \b} &=& \frac{2}{\sqrt{\gh}} [ \6_{( \a} \bar{c}_{\m}
 \6_{\b )} c^{\m}+
         \6_{\m} \bar{c}_{\a}  \6_{\b )}  c^{\m} +\6_{\m} \6_{( \a}
         \bar{c}_{\b )}  c^{\m} +
         \2 \6_{( \a} \bar{c}_{\b )} \gh_{\m \n} \6_{\l} \gh^{\m \n}
         c^{\l} - \nonumber \\
    & & \qquad - \2 \gh^{\m \n} \6_{\m} \bar{c}_{\n} \gh_{\r (\a} \gh_{\b )
    \s} \6_{\l} \gh^{\r \s} c^{\l}-
         \2  \6_{\l}( \gh_{\a \b} \gh^{\m \n} \6_{\m} \bar{c}^{\n}  c^{\l})
          ] \plabel{Tcc2} \\
T^{Bh}_{\a \b} &=&  \frac{1}{\sqrt{\gh}} \gh^{\m \n} \6_{\m} B_{\n}
h_{\a \b}
\ea
For the $B$-$h$ part one has
to take into account the tracelessnes of $h$.
The contributions to the anomaly coefficient for both
gauges are given in tab. 1 with $X, \bar{c}c$ and $Bh$
denoting the respective loop contributions.\\[.5cm]
\centerline{\begin{tabular}{c|c|c|c|c}
              gauge & $X$ & $\bar{c}c$ & $Bh$ & net result \\ \hline
              harmonic & $D$ & -52 & 26 & $D-26$   \\
              deDonder & $D$ & -28 & 2 & $D-26$  \\
\end{tabular}}\\[.3cm]
\centerline{{\small Tab1.: Anomly coefficient $c$ (eq. \pref{TTT})}}\\[.5cm]
These results are in agreement with those of
ref. \pcite{KR} obtained by calculating the effective action and
the result for the harmonic gauge is in agreement with ref. \pcite{Duese},
too. The general theorems on the gauge independence of
anomalies \pcite{Pigu} are thus seen to hold true.

\vspace{1cm}

{\bf Acknowledgements}

The author has profited from discussions with H. Balasin, W.Kummer,
H. Nachbagauer, A. Rebhan and
S. Sorella. He thanks W.Kummer and A. Rebhan for a critical reading of
the manuscript
and useful suggestions regarding the latter.

\end{document}